# Photocurrent Generation in a Metallic Transition Metal Dichalcogenide

Naveed Mehmood[1], Hamid Reza Rasouli[1], Onur Çakıroğlu[2], T. Serkan Kasırga[1,2*]

[1] UNAM- National Nanotechnology Research Center and Institute of Materials Science and Nanotechnology, Bilkent University, Ankara, Turkey 06800
[2] Department of Physics, Bilkent University, Ankara, Turkey 06800

* Corresponding Author E-mail: kasirga@unam.bilkent.edu.tr

**Abstract**

**Light induced current in two-dimensional (2D) layered materials emerges from mechanisms such as photothermoelectric effect, photovoltaic effect or nonlocal hot carrier transport. Semiconducting layered transition metal dichalcogenides have been studied extensively in recent years as the generation of current by light is a crucial process in optoelectronic and photovoltaic devices. However, photocurrent generation is unexpected in metallic 2D layered materials unless a photothermal mechanism is prevalent. Typically, high thermal conductivity and low absorption of the visible spectrum prevent photothermal current generation in metals. Here, we report photoresponse from two-terminal devices of mechanically exfoliated metallic 3R-$NbS_2$ thin crystals using scanning photocurrent microscopy (SPCM) both at zero and finite bias. SPCM measurements reveal that the photocurrent predominantly emerges from metal/$NbS_2$ junctions of the two-terminal device at zero bias. At finite biases, along with the photocurrent generated at metal/$NbS_2$ junctions, now a negative photoresponse from all over the $NbS_2$ crystal is evident. Among our results, we realized that the observed photocurrent can be explained by the local heating caused by the laser excitation. These findings show that $NbS_2$ is among a few metallic materials in which photocurrent generation is possible.**

**Text**

Photocurrent generation in semiconducting 2D layered materials is dominantly due to photothermal [1], photovoltaic effects [2,3] as well as excitation of nonlocal hot carriers [4-6]. In metals, these mechanisms typically do not result in photocurrent generation except in a few cases. Photothermal effects are generally not significant in metals because of typical high thermal conductivity and low absorption of the optical excitation. Moreover, optically excited electrons in metals would not have a measurable contribution to the large number of intrinsic electrons near the Fermi level when a bias is applied. In a few cases such as metallic carbon nanotubes [7], graphene [4,8,9] and gold nanoparticle networks [10], photocurrent generation has been reported. Light induced current generation in metallic 2D layered transition metal dichalcogenide (TMDC) is unprecedented, thus, it is not clear what mechanism will be prevalent. Here, we investigate photoresponse of mechanically exfoliated thin Niobium Disulfide ($NbS_2$) crystals using scanning photocurrent microscopy (SPCM) as an exemplary metallic TMDC.

$NbS_2$ can be found in layered form both in hexagonal (2H) and rhombohedral (3R) polytypes. While both polytypes are metallic [11], only 2H-$NbS_2$ is superconducting. It is reported in the literature [12-15] that 3R-$NbS_2$ shows a conductivity maximum, ranging from 20K to 50K, whose origin is still under debate [16,17]. The conductivity maximum may indicate charge density wave correlations [18] as well as weak localization [19], strong Coulomb correlations [20] or the Kondo effect [21]. Sketches of the 3R structure are depicted in Fig. 1(a) and 1(b).

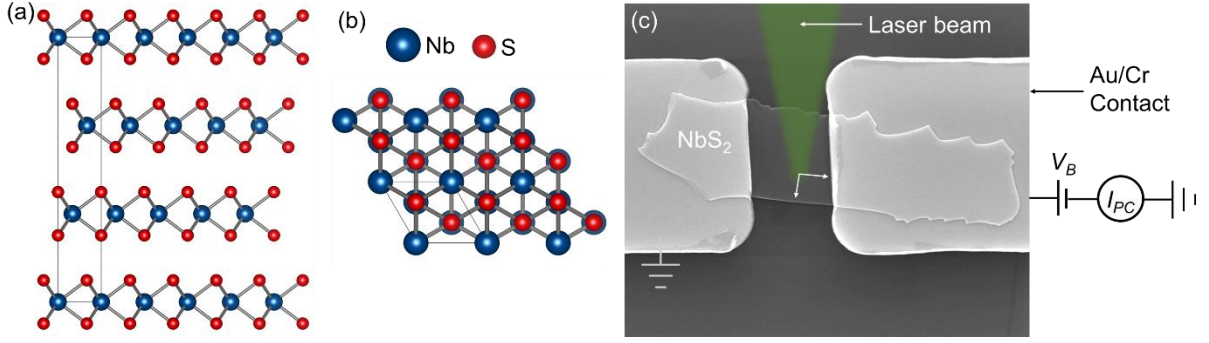

FIG. 1. (a) Crystal structure of 3R-NbS$_2$ is depicted from the side and (b) from the top. (c) Measurement configuration is depicted on a scanning electron microscope micrograph of a two-terminal device. Laser beam raster scans over the sample with a diffraction limited spot. Throughout the paper, left contact is grounded, and the bias is applied through the right contact using a current pre-amplifier.

We report photocurrent measurements on two terminal 3R-NbS$_2$ devices, far from the conductivity maximum point, near or at room temperature. NbS$_2$ flakes of various thicknesses are patterned using optical lithography followed by gold/chromium electrode deposition. These devices will be referred as top contact (TC) devices. Figure 1(c) depicts the scanning laser beam as well as the electrical connection on a scanning electron microscope (SEM) micrograph of a TC device. Same electrical connection configuration is used throughout the paper, ground terminal is connected to the left, current pre-amplifier terminal is connected to the right contact (See Supplementary Information for further details).

Optical image of a typical TC device (named TC-1) is shown in Fig. 2(a). Reflection and photocurrent maps obtained with a 532 nm laser of power $P = 85$ µW (~11 kW cm$^{-2}$) focused to a diffraction limited spot are given in Fig. 2(b) and 2(c) respectively. Resistance vs. temperature (R-T) measurement given in Fig. 2(d) from TC-1 shows a ~21.5 Ω resistance (R) at room temperature. R-T measurement clearly shows the metallic nature of the NbS$_2$ flake. Atomic force microscopy (AFM) height trace gives the flake thickness as ~180 nm (AFM measurement is provided in Supplementary Information). The photocurrent map for TC-1 under zero bias reveals that the extremum current is generated at the junctions where NbS$_2$ flake meets with the metallic contacts. We consider that the photocurrent in this device has a photothermal origin as the other mechanisms are less likely in metal/metal junctions.

Firstly, we study the SPCM results on TC-1 device in more detail (Fig. 2(a)-(e)). Due to the Seebeck effect, the lattice temperature difference $\Delta T_C$ between the two terminals of the contacts will generate a thermoelectric electromotive force (emf), $V_T = -\Delta S_{Au/NbS_2} \Delta T_C$. Using the absolute Seebeck coefficients reported in the literature for NbS$_2$ and Au, -4 µV/K and ~2 µV/K respectively [16,22], the difference in the Seebeck coefficients of NbS$_2$ and Au is $\Delta S_{Au/NbS_2} = S_{Au} - S_{NbS_2} \approx -6$ µV/K. When the laser is focused at the junction, the maximum emf generated is $V_{PC} = RI_{PC} \approx 4.5$ µV, where $I_{PC}$ is the measured photocurrent. $I_{PC}$ measured on the right and the left junctions are -130 and 150 nA respectively. Comparing these $V_{PC}$ values to $V_T$ shows that $\Delta T_C \approx 0.54$ K ($\approx 0.46$ K for the left junction) for $P = 85$ µW, which corresponds to 6.3 mK/µW (5.5 mK/µW for the left junction). These values of $\Delta T_C / P$ are consistent with the values reported in similar studies in the literature [23,24].

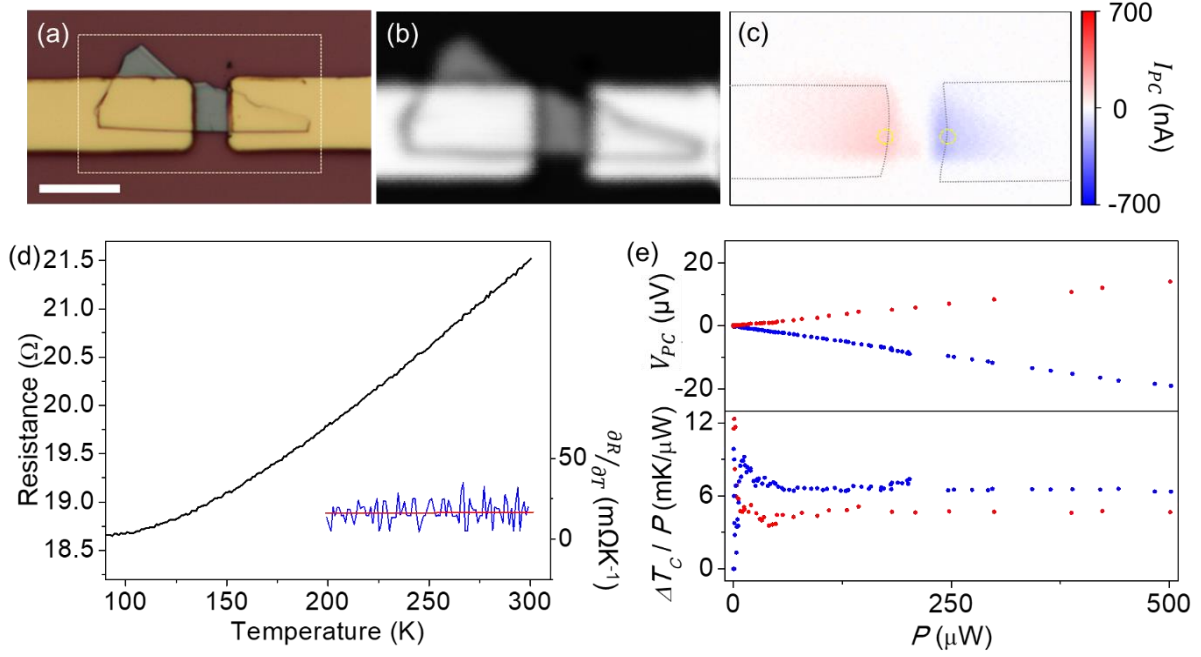

FIG. 2. (a) Optical microscope image of TC-1 is shown. SPCM measurement is taken from the region boxed with a dashed rectangle. Scale bar is 10 μm. (b) Reflection map and (c) the corresponding photocurrent map of TC-1 under zero-bias shows photocurrent emerging around the contacts. Black dashed lines indicate the outlines of the contacts. (d) *R-T* graph shows the metallic characteristic of the sample. $\partial R / \partial T$ is given in the inset from 200 to 295 K as a reference. (e) Upper panel shows $V_{PC}$ vs $P$ measured from the metal/NbS$_2$ junctions (positions indicated by yellow dashed circles in (c)), red points from the left contact and blue points from the right contact. Lower panel shows the calculated $\Delta T_C / P$ from the $V_T$ vs $P$ data.

To test if we get a consistent behavior for different $P$ values, we parked the laser spot on the metal/NbS$_2$ junctions where the photocurrent is at its maximum and minimum, marked by yellow dashed circles in Fig. 2(c). Then, while simultaneously reading the laser power through a 50:50 beam splitter, we tuned the laser power with a variable neutral density filter and recorded the photocurrent. Figure 2(e) upper panel shows $V_{PC}$ vs $P$. For both junctions we observe a similar linear change in the photoresponse with increasing laser power. Moreover, when we plot the calculated temperature increase per unit laser power vs. the laser power (Fig. 2(e) lower panel), we see that at both junctions, $\Delta T_C / P$ has a similar value with a small difference. This difference in $\Delta T_C / P$ at the opposing junctions might be due to slight differences in the gold contact edges (Fig. 1(c)) as well as positioning of the laser spot. NbS$_2$ crystals transferred on top of prepatterned gold contacts also exhibit a similar photoresponse to the TC devices as discussed later in the text.

When we apply bias to the device, while the photoresponse at the metal/NbS$_2$ junction changes, we now observe a photoresponse all over the crystal. Figure 3(a) and 3(b) show photocurrent maps of TC-1 taken under 50 mV and -50 mV biases ($V_B$), and Fig. 3(c) upper panel shows the photoresponse along the center of the device at several finite biases. We calculate the photoconductance $G_{PC} = (I_B - I_0)/V_B$ where $I_0$ is the zero-bias photocurrent and $I_B$ is the photocurrent at any given $V_B$. $G_{PC}$ values for various biases are the same throughout the crystal (Fig. 3(c) lower panel). The change in photoresponse due to the applied bias indicates a bias-independent photoconductance mechanism.

The photoresponse observed from all over the crystal under bias can be explained by the local temperature increase, $\delta T_L$, caused by the laser. As the temperature increases locally, the electrical conductivity decreases on and around where the laser spot hits. This decrease leads to a negative photoconductance in such a way that the dc current due to the applied bias decreases. We calculated the change in the resistance of the device with the approximation that the temperature increase within a disk of diameter $D \approx 1$ μm is uniform and outside the disk, it is zero (see Supplementary Information for details). For a device like TC-1, this calculation shows that for a laser induced local temperature rise of $\delta T_L / P \approx 40$ mK/μW the magnitude of the measured photocurrent is in excellent agreement with the calculated value. This local temperature change being slightly higher than the temperature increase extracted from the metal/NbS$_2$ junction, $\Delta T_C / P$, is consistent with the lower thermal conductivity of SiO$_2$ substrate as compared to Au/Cr.

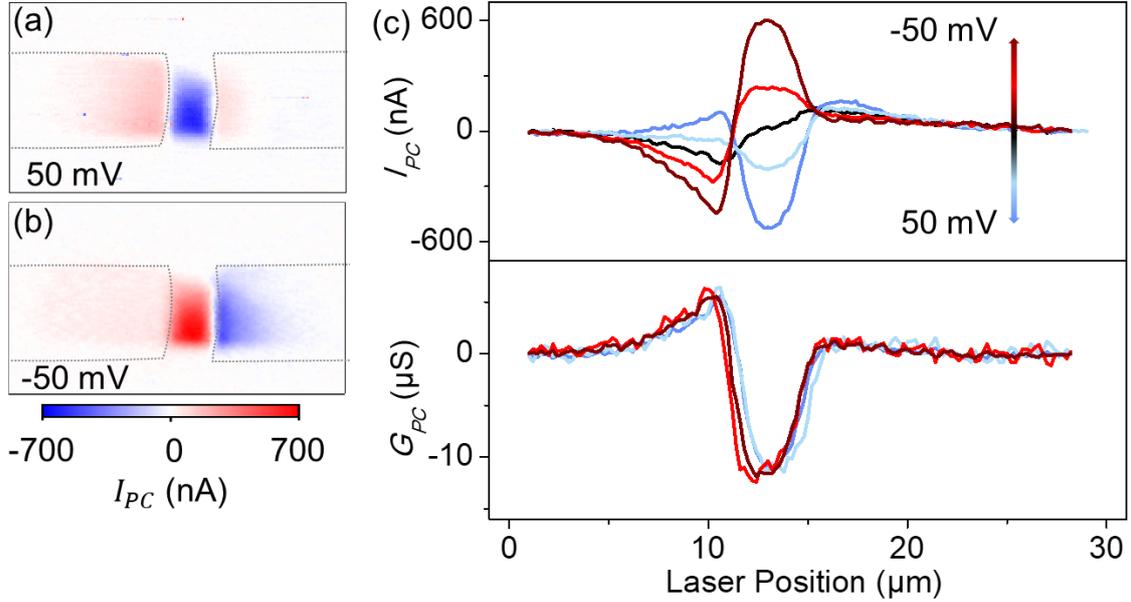

FIG. 3. (a) Photocurrent map of TC-1 under 50 mV and (b)-50 mV bias is given. Photocurrent is generated all over the crystal between the contacts when the bias is applied. Dashed gray line represents the outline of the metal contacts. (c) Line trace along the center of the crystal shows the photocurrent (upper panel) under -50, -20, 0, 20 and 50 mV biases and respective photoconductances (lower panel). It is clear that the photoconductance $G_{PC}$ is the same for all different biases.

Negative photoconductance observed from the center of the crystal should diminish in a device with poor electrical contacts as the effective bias on the crystal ($V_{Eff}$) will be much lower due to the potential drop through the contact resistance, $R_C$. This is what we observe in crystals transferred on top of pre-patterned thin metallic contacts. Figure 4(a) shows a bottom contact device (BC-1) with the NbS$_2$ crystal (~100 nm thick) transferred using the polycarbonate transfer method [25] on top of 10/5 nm thick Au/Cr contacts. The resistance of BC-1 is measured as $R = 600$ Ω at room temperature. Firstly, zero-bias photocurrent map obtained with a 532 nm laser of power $P = 32$ μW (Fig. 4(b)) shows a local photoresponse unlike TC devices. The magnitude of the zero-bias photoresponse is an order of magnitude smaller. This can be attributed to the higher resistance of the device. When we calculate the $\Delta T_C / P$ value for BC-1 we get about 50 mK/ μW which is higher as compared to TC devices. This is also expected as the thermal contact of NbS$_2$ flake with the contact pads is poorer compared to TC devices.

Based on our resistivity measurements on TC-1, a crystal like the one used in BC-1 should have a resistance of ~25 Ω at room temperature. The difference between the expected resistance, $R_E$, and the measured $R$ can be attributed to the contact resistance. The photoresponse at zero bias is localized to small regions on the contacts and the strength of the photothermal current is different in amplitude for the ground and the pre-amp side. This is an indicator of varying contact quality across the contacts. When a finite bias is applied, only the signal coming from the NbS$_2$/metal junctions is altered and we do not observe a photoresponse throughout the crystal (Fig. 4(c)-(d)). $V_{Eff}$ on the crystal can be written as $V_{Eff} = \frac{V_B}{R_E + R_C} R_E$. For BC-1, $V_{Eff}$ is as low as 4% of $V_B$. Thus, photocurrent coming from the center of the crystal is less than a nanoampere in a device like BC-1.

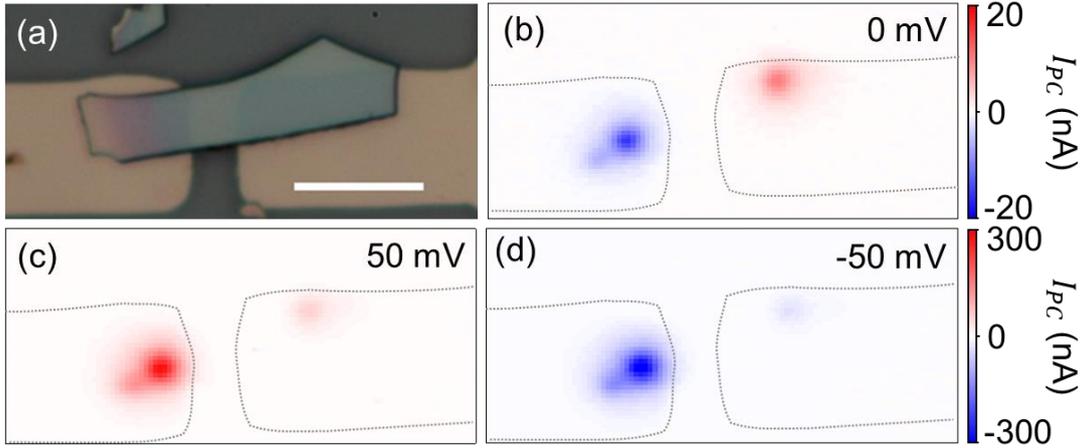

FIG. 4. (a) Optical microscope image of bottom contact device (BC-1) is shown. Scale bar is 10 μm. Photocurrent maps (b) under zero-bias, (c) 50 mV and (d) -50 mV bias show that there is no observable photoconductance change in the center of the crystal. This can be explained by the fact that the contact resistance of BC-1 is so high that $V_{Eff}$ on the crystal is small and hence local resistance change caused by photothermal heating becomes insignificant. Dashed gray line represents the outline of the metal contacts.

The results we report in this work demonstrate for the first time that the light induced current generation even in metallic TMDC is possible. SPCM measurements reveal the photothermal origin of the generated photocurrent and this result is consistent with the correlated nature of 3R-NbS$_2$ crystals [26]. We would like to note that we performed SPCM experiments under vacuum to rule out any other possible photocurrent mechanisms due to oxygen in the ambient [27]. We observe no difference between the photoresponse of devices under vacuum or in ambient (Supplementary Information). We also performed SPCM on indium contact devices from which photoresponse similar to TC devices has been observed (see Supplementary Information for details). This study shows that 3R-NbS$_2$ sets an example to a few other metallic materials that photocurrent generation is possible. Our findings will be useful in engineering of all-TMDC optoelectronic components [28].

**Acknowledgements**

We would like to thank Engin Can Sürmeli, Koray Yavuz, Breera Maqbool and Talha Masood Khan for their useful comments on the manuscript. This work is supported under TUBITAK 1001 program, grant no: 214M109.

**Additional Information**

Supplementary information is available.

# Photocurrent Generation in a Metallic Transition Metal Dichalcogenide– Supplementary Information

Naveed Mehmood[1], Hamid Reza Rasouli[1], Onur Çakıroğlu[2], T. Serkan Kasırga[1,2,*]

[1] UNAM- National Nanotechnology Research Center and Institute of Materials Science and Nanotechnology, Bilkent University, Ankara, Turkey 06800
[2] Department of Physics, Bilkent University, Ankara, Turkey 06800

* Corresponding Author E-mail: kasirga@unam.bilkent.edu.tr

**Experimental Methods**

**Device Fabrication**

$3R-NbS_2$ flakes are exfoliated on to a 280 nm $SiO_2$/Si substrate from a commercially available (HQ Graphene, the Netherlands) bulk crystal using the Scotch tape method [1]. The flakes are characterized using selected area electron diffraction (SAED), Raman and energy dispersive X-ray spectroscopy (EDX). After finding thin flakes under optical microscope by judging from the optical contrast, their thicknesses are measured using AFM. Top contact devices are fabricated using standard optical lithography followed by 100/10 nm Au/Cr evaporation. For the bottom contact devices, $NbS_2$ crystals are transferred on to pre-patterned Au/Cr contacts using the polycarbonate (PC) transfer method [2]. First, the $NbS_2$ flake under study is picked-up from the exfoliated surface by melting a $mm^2$ piece of PC on the crystal. Once it cools down, the PC film is gently peeled off. This picks up the flake. Then, under optical microscope, the flake is aligned with the gold contacts using a micromanipulator. After re-melting the PC film, it is dissolved in chloroform while leaving the $NbS_2$ flake on the contacts. Indium contacted devices reported in Supplementary Information are made by drawing fine indium pins from a molten indium blob and placing these pins on to the heated $NbS_2$ flake using a micromanipulator. When the solidified indium pin touches to the hot $NbS_2$ flake, it re-melts and contacts the flake.

**Scanning Photocurrent Microscopy Setup**

We use a home-brew scanning photocurrent microscopy setup. A diffraction limited laser beam chopped at a certain frequency (typically at 1.4 kHz) raster scans the sample. Via electrical probes, one contact is grounded while the other contact is connected to a current pre-amplifier (SRS SR570) whose output is fed into a lock-in amplifier (SRS SR830) to read the laser induced photocurrent. The reflected beam is collected to form the reflection map via a silicon detector whose output is also detected with a lock-in amplifier. 532 nm laser diode is used as the excitation source.

**Raman, EDX Spectrum and, Selected Area Electron Diffraction of $3R-NbS_2$**

We performed Raman measurements on most of the samples to confirm that the crystals we exfoliate are $3R-NbS_2$. Also on a few samples we gathered selected area electron diffraction (SAED) using transmission electron microscope (TEM). Raman spectrum of a sample is given in Figure S1a. Both A and E peaks are labelled on the graph and their positions are in agreement with the values reported for $3R-NbS_2$ in the literature[3].

TEM image of a flake transferred on a TEM grid is given in Figure S1b and the SAED pattern is given in Figure S1c. In plane lattice constant (*a*) calculated from the diffraction spots reveals 0.333 nm which is in excellent agreement with the literature[4]. EDX maps also confirm that we have Nb an S and intensity peaks from several points confirm Nb:S ratio to be around 1:2.

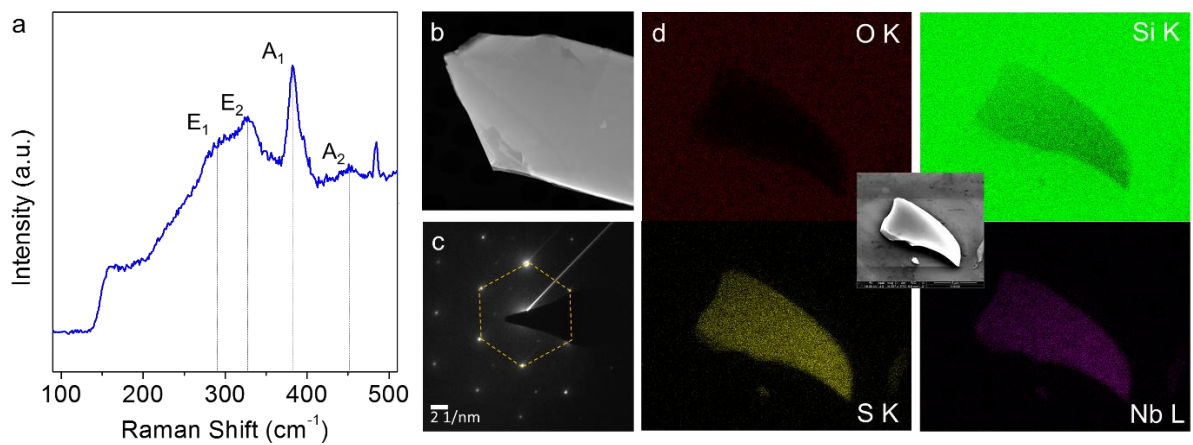

**Figure S1 a.** Raman spectrum shows the characteristic peaks corresponding to the 3R-$NbS_2$. Relevant peaks are labelled. **b.** TEM image of a typical exfoliated crystal and **c.** SAED image is given. The separation of the diffractions spots match with the inter-planar spacing reported in the literature. **d.** EDX maps of O K, Si K, S K and Nb L. Inner SEM micrograph shows the sample we scanned.

### Photothermal Photoconductance Calculations

As mentioned in the main text, to calculate the local temperature rise caused by the laser spot on the crystal we assumed that the temperature increase $\delta T_L$ on a disk of diameter of 1 μm is uniform and outside the disk, it is zero. This is not a bad approximation as the laser spot is a diffraction limited spot with a Gaussian profile.

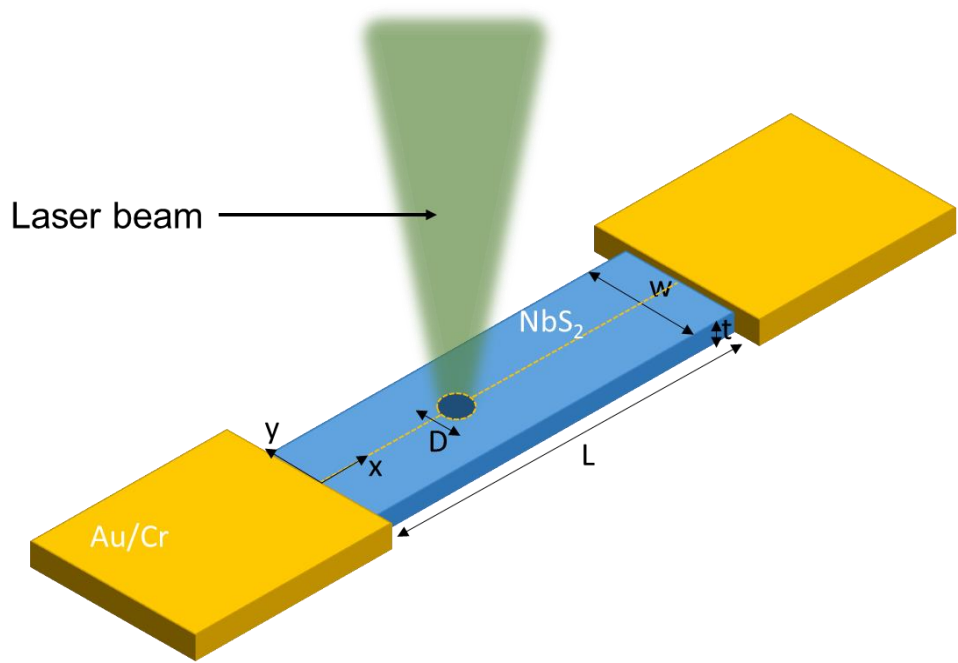

**Figure S2 | Schematic of the device parameters.** Schematic shows the parameters used in the calculation of $\delta T_L$.

For the ease of calculations, we assumed the laser spot is positioned at the center of the crystal at an arbitrary distance $x$ from the contacts. The coordinates for the center of the laser beam is $(x, 0)$. The radius of the laser beam temperature rise disk is $D$. Within this radius the resistivity of the crystal will be different then the rest of the crystal. If we take an infinitesimally thin slice of crystal going through the laser beam temperature rise disk across the contacts, resistance of the slice can be written as:

$$dR = \rho_0 \frac{x_i}{A} + \rho_0 \frac{L - x_f}{A} + \rho_T \frac{x_f - x_i}{A}$$

where $\rho_0$ is the resistivity of the crystal at ambient temperature, $\rho_T$ is the resistivity under the laser spot, $L$ is the total length across the contacts, $A$ is the cross-sectional area of the crystal and, $x_f = D + \sqrt{D^2 - y^2}$ and $x_i = D - \sqrt{D^2 - y^2}$ are the edges of the laser beam temperature rise disk along the x axis. To determine the value of $\rho_T$, based on the $R - T$ measurements and the device dimensions we find the slope, $\frac{\partial \rho_T}{\partial T}$ at room temperature and write $\rho_T = \delta T_L \frac{\partial \rho_T}{\partial T} + \rho_0$. The total conductance of the crystal with laser spot on can be written as the sum of all parallelly connected infinitesimally small stripes of resistances:

$$\frac{1}{R_T} = \frac{y_1 t}{\rho_0 L} + \frac{(w - y_2)t}{\rho_0 L} + \int_{y_1}^{y_2} \frac{t dy}{\rho_0 (L - (x_f - x_i)) + \rho_T (x_f - x_i)}$$

Here $w$ is the total width and $t$ is the thickness of the crystal. Evaluating the integral from $y_1 = 0$ to $y_2 = D/2$ and multiplying the result by 2 will give the same result as evaluating it from $y_1 = -D/2$ to $y_2 = D/2$.

$$\frac{2}{R_T} = \left[ \frac{\sigma \pi/4 - \sqrt{2}b \tan^{-1}(\sqrt{2}\frac{b}{\sigma}) + \sqrt{2}b \tan^{-1}(\frac{a.D}{\sqrt{2}\sigma})}{a.\sigma} + \frac{1}{2}\frac{w - D}{\rho_0 L} \right].t$$

where $a = 2(\rho_T - \rho_0)$, $b = \rho_0 L$, $\sigma = (2b^2 - a^2 D^2)^{1/2}$. Based on the equation we derive above, we wrote a Python code that can calculate $I_{PC}$ based on given device parameters such as dimensions and resistance. We tune $\delta T_L$ value to find the measured $I_{PC}$.

**Low Pressure Scanning Photocurrent Measurements and BC Device with Low Contact Resistance**

To exclude any possible effects due to oxygen in the ambient, we performed scanning photocurrent measurements at about 5 10$^{-2}$ mBar on a TC device, TC-3 as well as on a BC device, BC-2 (Figure S3). Different scans taken at various biases show no difference between the ambient and low pressure photoconductances of TC-3. For BC-2 photocurrent maps taken at various biases show no difference in ambient and low pressure.

Another observation we would like to report on BC-2 is the observed photoconductance from all over the crystal. As mentioned in the main text, as the contact resistance increases the resistance change induced by the laser heating becomes insignificant. BC-2 is a device made with a thinner (~30 nm) and a larger flake so the contact area with the bottom pads are much larger. This leads to a more significant photoconductance that can be clearly seen in Figure S3b-c. Resistance of BC-2 is measured as 36 Ω both in ambient and vacuum (Figure S3e).

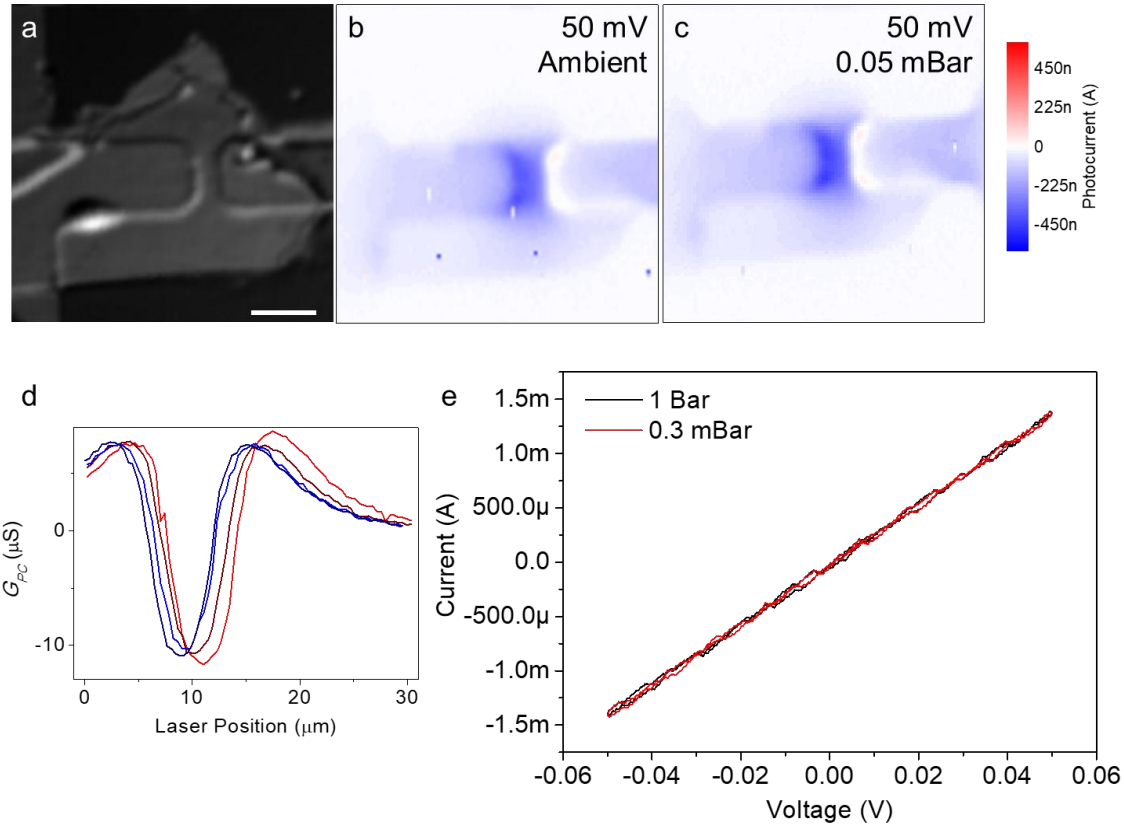

**Figure S3 a.** Reflection map of BC-2. Scale bar is 10 μm. **b.** and **c.** shows SPCM map for BC-2 taken under 50 mV in ambient and at 0.05 mBar respectively. There is no difference between low pressure and ambient photoresponse. **d.** Low pressure photoconductance trace taken through TC-3 device shows no difference between low pressure (blue curves) and ambient (red curves) taken at 50 and -50 mV biases. **e.** IV curves for BC-2 taken under ambient and 0.3 mBar shows no difference except a shift that is due to the centering of the line trace.

## Laser Power Dependence of Photoconductance

One test we performed on TC devices is the laser power dependent photoconductance from the crystal. Figure S4 shows TC-4 and photocurrent maps at 50 and -50 mV biases. As in other samples the photoconductance from the bulk of the crystal is negative and shows a linear power dependence. When we calculate the photoconductance per unit laser power we see that the spread of values is very small with a mean value of -0.068 S/W. When we calculate the laser induced heating around the center using the method explained earlier we find 35 mK/μW. This value is similar to the value we report for TC-1 in the main text.

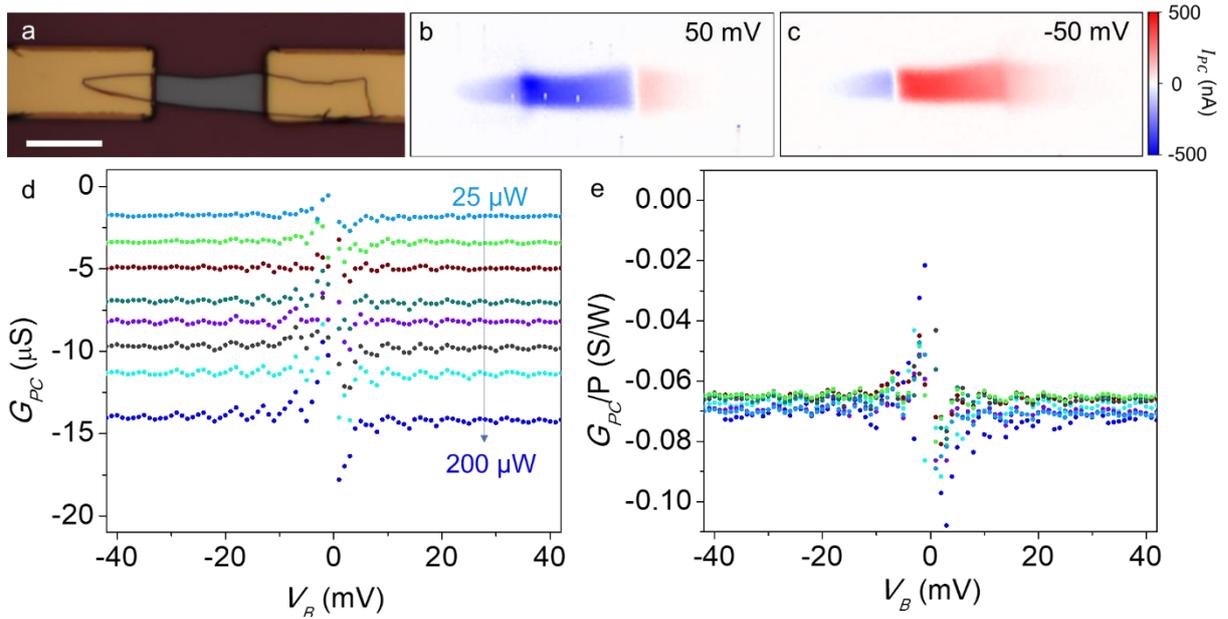

**Figure S4 a.** Optical microscope image of TC-4. Scale bar is 10 µm. **b.** Photocurrent map under 50 mV and **c.** -50 mV bias. **d.** Laser power dependence of photoconductance for various biases from the center of the crystal is given for laser powers from 25 µW to 200 µW with 25 µW increments. **e.** Photoconductance per unit laser power has a mean value of -0.068 S/W for various biases.

### Indium Contact Device

To test whether the contact metal makes any difference in photoresponse, we placed indium contacts on $NbS_2$ flakes. First, we draw an indium pin from a molten blob of indium using a micromanipulator. Then, the indium pin is placed on heated $NbS_2$ surface. Once the sample cools down, indium solidifies and makes an electrical contact with the flake. Optical image of an indium contact device (ID-1) is given in Figure S5. Resistance of the crystal is measured to be 196 Ω.

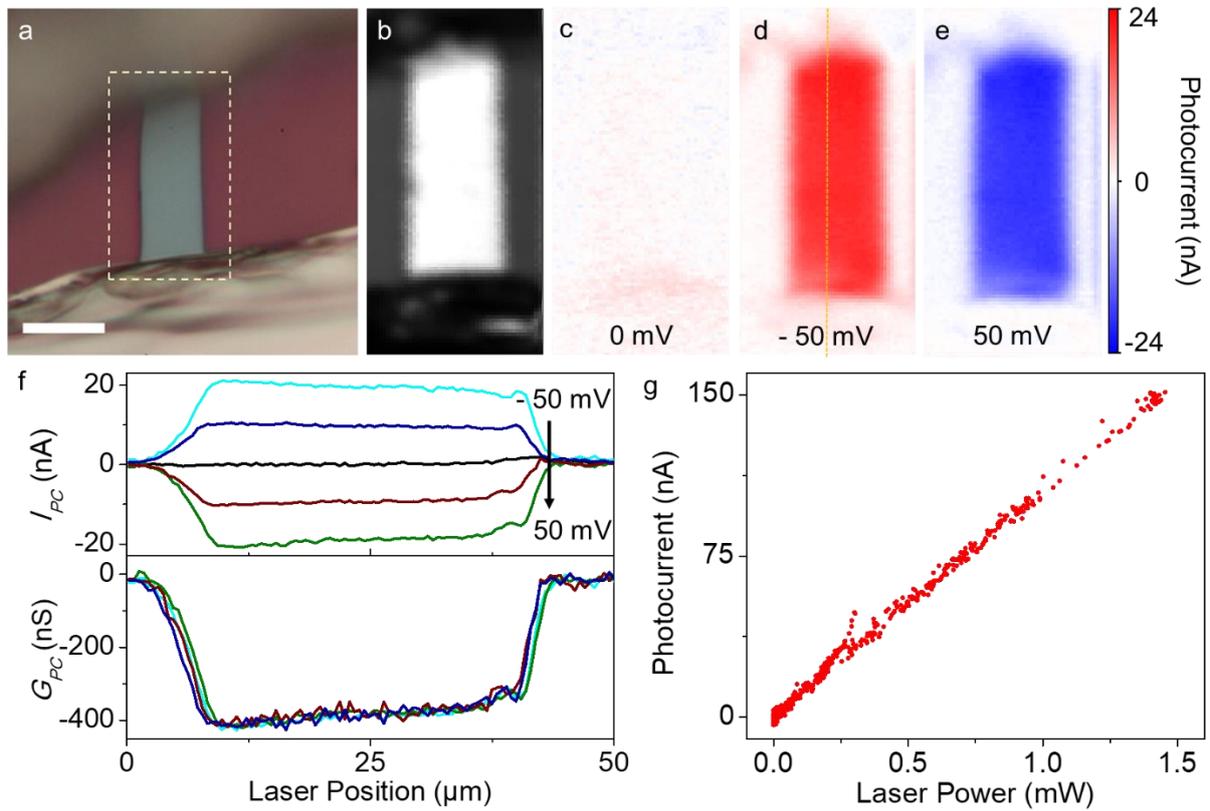

**Figure S5 a.** Optical microscope image of ID-1. White dashed rectangle shows the region where SPCM is taken. Scale bar is 10 μm. **b.** Reflection map and corresponding photocurrent maps at **c.** 0 mV **d.** -50 mV **e.** 50 mV biases. Dashed line through the center of the crystal shown in **d** denotes the line traces given in **f** for various biases. Although the Seebeck contribution is now small, photoconductance is the same for all biases. **g.** Laser power dependence of the photocurrent collected from the center of crystal shows a linear dependence.

Most striking difference in SPCM between TC and ID devices is the weakness of the Seebeck signal in zero bias as can be seen in Figure S5c. This can be explained by the fact that the laser beam is blocked by very thick indium pins, thus heating of the In/NbS$_2$ junction is miniscule. However, when we apply bias (Figure S5 d,e,f), we observe a negative photoconductance similar to TC devices. Moreover, laser power dependence of the photocurrent from the center of the crystal is very similar to TC devices.

**Top Contact Device-1 (TC-1)**

3R-NbS$_2$ flakes are exfoliated on a silicon chip with 280 nm oxide layer on top. AFM measurement in Figure S6 shows a slight slope on the crystal for the height trace. We took the average height as the height of the crystal in our calculations.

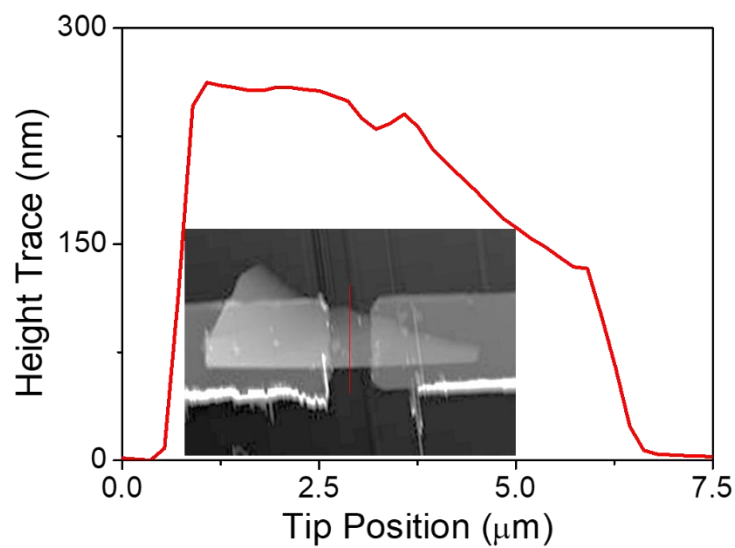

**Figure S6** AFM height trace taken along the red line indicated in the inset shows there is a slight slope in the crystal that could be coming from crystal itself.